# METHODS OF PHASE AND POWER CONTROL IN MAGNETRON TRANSMITTERS FOR SUPERCONDUCTING ACCELERATORS*

G. Kazakevich#, R. Johnson, M. Neubauer, Muons, Inc, Batavia, IL 60510, USA
V. Lebedev, W. Schappert, V. Yakovlev, Fermilab, Batavia, IL 60510, USA

*Abstract*

Various methods of phase and power control in magnetron RF sources of superconducting accelerators intended for ADS-class projects were recently developed and studied with conventional 2.45 GHz, 1 kW, CW magnetrons operating in pulsed and CW regimes. Magnetron transmitters excited by a resonant (injection-locking) phase-modulated signal can provide phase and power control with the rates required for precise stabilization of phase and amplitude of the accelerating field in Superconducting RF (SRF) cavities of the intensity-frontier accelerators. An innovative technique that can significantly increase the magnetron transmitter efficiency at the wide-range power control required for superconducting accelerators was developed and verified with the 2.45 GHz magnetrons operating in CW and pulsed regimes. High efficiency magnetron transmitters of this type can significantly reduce the capital and operation costs of the ADS-class accelerator projects.

## INTRODUCTION

Accelerating cavities of the modern superconducting CW accelerators operate with very high loaded quality factors, $Q_L$. The detuning caused by beam loading, microphonics, and other external sources, can be as large or larger than the bandwidth of the accelerating mode [1]. Detuning results in parasitic amplitude and phase modulations of the accelerating field independent in each SRF cavity. The RF source must have sufficient rate dynamic control of the amplitude and phase of the RF field in each SRF cavity, [2] to maintain a stable field in the presence of such detuning. Traditional RF sources (klystrons, IOTs, solid-state amplifiers) can provide such control but at a cost. The capital cost of the RF system of the superconducting accelerator will be a significant part of the project cost. Magnetrons controlled by a resonant RF signal may provide the required dynamic phase and power control bandwidths at a significantly lower the capital cost, at a higher efficiency, [3]. Methods of phase and power control magnetrons are discussed.

## A WIDEBAND PHASE CONTROL IN MAGNETRON TRANSMITTERS

Phase modulation in injection-locked magnetron transmitters was studied in Ref. [4] using 2.45 GHz, 1 kW magnetrons operating with 5ms pulses. Both single-cascade and 2-cascade magnetron were examined using the setup shown schematically in Fig 1. The 2-cascade configuration provides the same bandwidth of the phase control at lower power of the injection-locking signal.

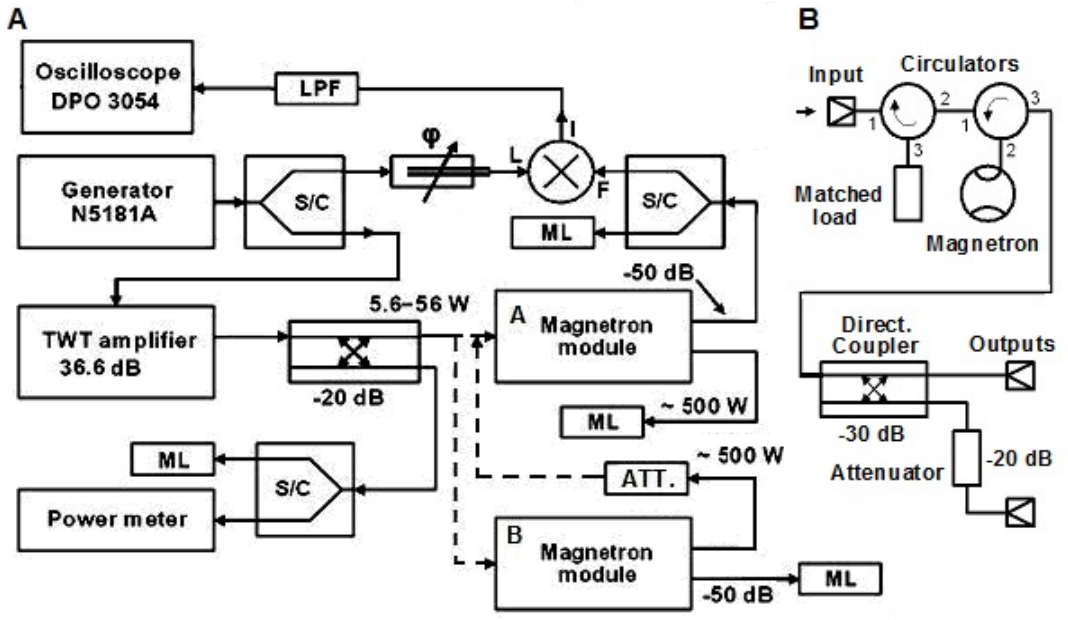

Figure 1: A- schematics for test of single and 2-cascade magnetrons by a phase-modulated frequency-locking signal. S/C is a splitter/combiner, LPF is a low pass filter, ML is a dummy load, and ATT is an attenuator. B- schematics of the magnetron module.

The single injection-locked magnetrons were tested in a configuration using module A with the magnetron injection-locked by a CW TWT amplifier and fed by a High Voltage (HV) modulator. Module B was disconnected from the amplifier and the modulator. The 2-cascade magnetron was tested in a configuration in which the magnetron in module B was injection-locked by the TWT amplifier while the magnetron in module A was connected via an attenuator to the output of module B. Both magnetrons were fed by the same modulator. Experiments demonstrated operation of the 2-cascade magnetron in injection-locked mode for the attenuator values in the range of 9-20 dB, [4].

An N5181A generator produced the wideband phase modulation of injection-locking signal for both the single and 2-cascade magnetrons. The broadband (2-4 GHz) TWT amplifier did not distort the phase-modulated injection-locking signal.

The transfer function magnitude characteristics averaged over 8 pulses of the single and 2-cascade magnetrons, Fig. 2, were measured in the phase modulation domain by the Agilent MXAN9020A Signal Analyzer over a range of injection-locking signal powers, $P_{Lock}$.

The measurements for the single and 2-cascade cascade magnetrons were performed with a phase modulation magnitude of 0.07 rad., and a magnetrons output power of $P_{Out}$ ≈450 W, [4]. The measurements with the 2-cascade magnetron were carried out with an attenuator value of ≈13 dB, *i.e.*, the second cascade (tube) operated at $P_{Lock}$ ≈25 W. In Figs. 2 and 3 in the plots concerning the 2-cascade magnetron, $P_{Lock}$ denotes power of the signal injection-locking the first cascade.

* Supported by Fermi Research Alliance, LLC under Contract No. De-AC02- 07CH11359 with the United States DOE in collaboration with Muons, Inc.
#e-mail: gkazakevitch@yahoo.com; grigory@muonsinc.com

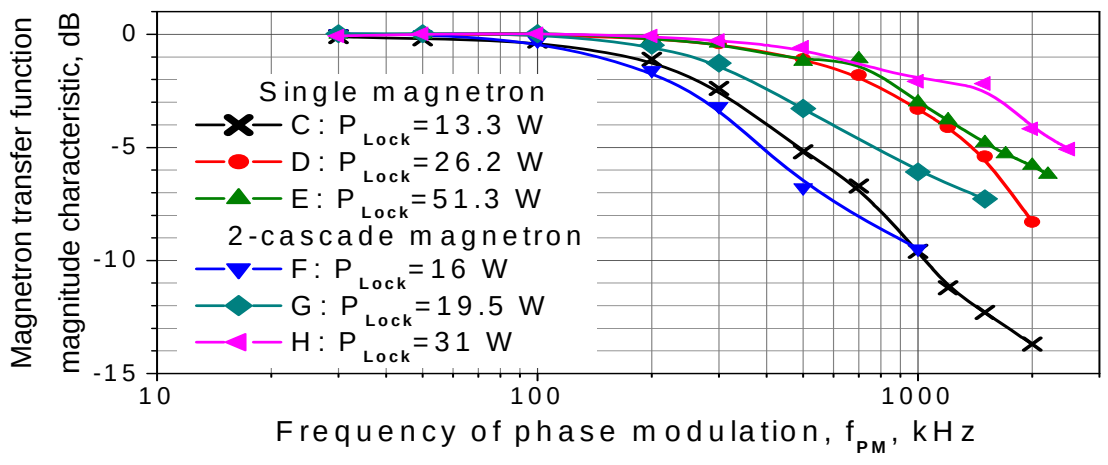

Figure 2: Transfer function magnitude characteristics of the phase control measured in the phase modulation domain with single and 2-cascade magnetrons in dependence on power of the injection-locking signal, $P_{Lock}$.

The phase response of the injection-locked magnetron to the modulation frequency, $f_{PM}$, has been measured with a calibrated phase detector including phase shifter φ, double balanced mixer and Low Pass Filter, LPF, at a modulation magnitude of 0.35 rad, and the magnetron output power of $P_{Out}$ ≈500 W, [4]. The transfer function phase characteristics considering the measured phase response, transfer function magnitude characteristics and the phase detector instrumental function are plotted in Fig. 3 for single and 2-cascade phase-controlled magnetrons over a range of injection-locking power.

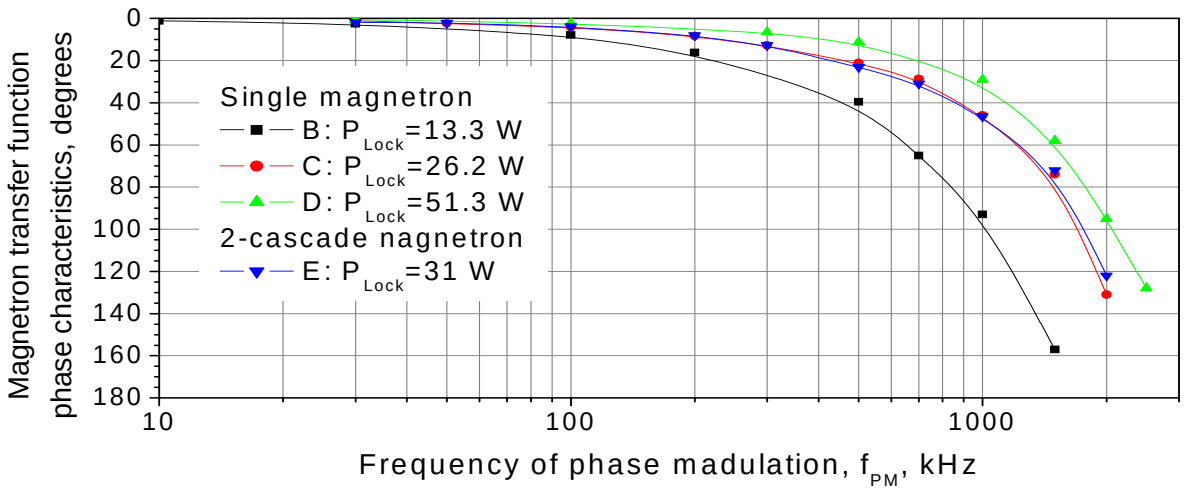

Figure 3: Transfer function phase characteristics of the single and 2-cascade magnetrons injection-locked by the phase-modulated signal vs. the modulating frequency, $f_{PM}$.

## A FAST POWER CONTROL IN MAGNETRON TRANSMITTERS

It has been shown that wideband phase modulation can also be used to modulate the power of a magnetron transmitter via two methods of vector power control.

The first method of the fast power control, [4], was realized in a two-channel transmitter with a combining of the output signals by a 3-dB hybrid combiner. The method allows any value of $Q_L$. The phase control of the combined signal provides a simultaneous phase control in the channels, while for power control was used a phase management with a controlled phase difference.

The second method requires only a single-channel magnetron transmitter, [5]. The phase-modulated injection-locking signal provided a fast phase control. Power control was provided by an additional modulation of the depth of the phase modulation. Such a modulation results in the RF power distribution between the fundamental frequency and the sidebands. If the frequency of the depth modulation is much larger than the accelerating field bandwidth, the power concentrated in the sidebands is reflected from the cavity into the dummy load. Thus, changes of the modulation depth result in a power change at the fundamental frequency. Note that this method of vector power control well operates at $Q_L \geq 10^7$.

Since the RF sources intended for powering of the SRF cavities using both vector methods of power control operate at a nominal power and a part of the power is continuously redistributed into the dummy load for absorption, both vector methods provide an average efficiency about of 50%-70% in dependence on the range of power control (a few dB) as it is required for SRF cavities.

A third method proposed and studied recently, [6], provides a significantly higher average efficiency (more than 80%) via a single magnetron with a range of power control up to 10 dB allowing a simultaneous wideband phase control. The power control in this technique is realized by a control of the magnetron current over an extended range when the magnetron driven by a sufficient resonant RF signal may operate below the threshold for self-excitation. The bandwidth of the power control in this technique is determined by the bandwidth of the current feedback loop in the magnetron HV power supply. Presently the bandwidth may be up to 10 kHz without compromising the efficiency of the HV power supply.

The feasibility of this method was demonstrated using 2.45 GHz, 1kW magnetrons operating in pulsed and CW regimes, [6]. A detailed study of the method was performed in CW regime with the 1.2 kW magnetron type 2M137-IL with a permanent magnet in setup resembling one shown schematically in Fig. 1B. The magnetron was fed by HV switching power supply type SM445G with a narrow bandwidth current feedback loop. The magnetron was driven by HP 8341A generator with solid-state and TWT amplifiers providing power of the injected resonant signal up to 100 W. Measured V-I characteristic of the magnetron at $P_{Lock}$=100 W is shown in Fig. 4.

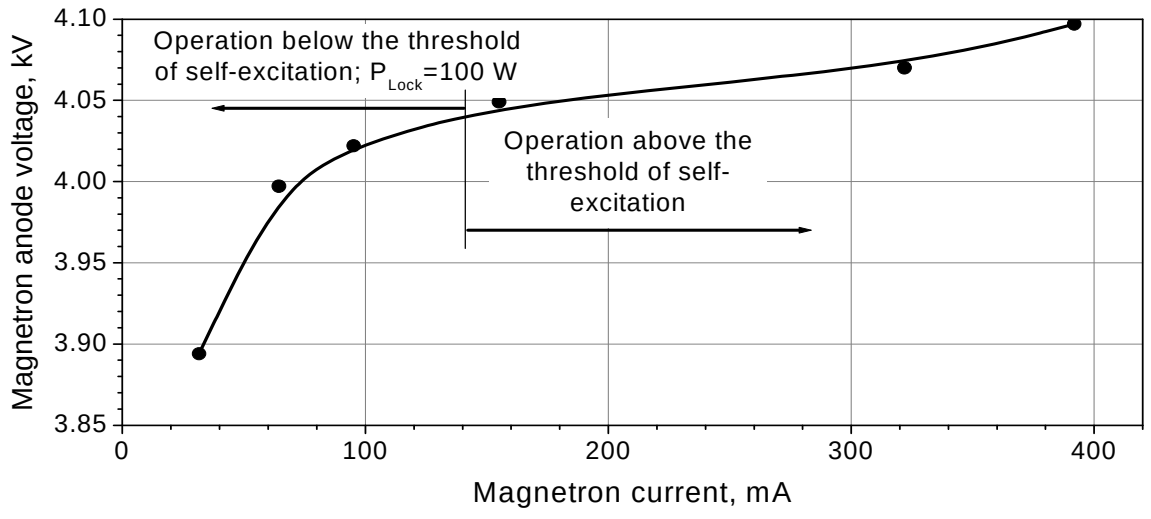

Figure 4: Measured magnetron V-I characteristic. The solid line (B-spline fit) shows an extended range of current with stable operation of the tube at the given $P_{Lock}$.

Fig. 4 indicates stable operation of the magnetron over 10 dB power range (0.1 kW-1.2 kW) which corresponds to the magnetron controlled current range of (31.2-392 mA) at $P_{Lock}$ =100 W. In accordance with Fig. 4 at the magnetron current <140 mA the magnetron voltage is less then the Hartree voltage and the magnetron starts up and stably operates due to compensation of a lack of magnitude of the synchronous wave by the injected resonant RF signal, [6]. Over the range of power of 10 dB the magnetron demonstrates precisely-stable carrier frequency at low noise (<-50 dBc), Fig. 5.

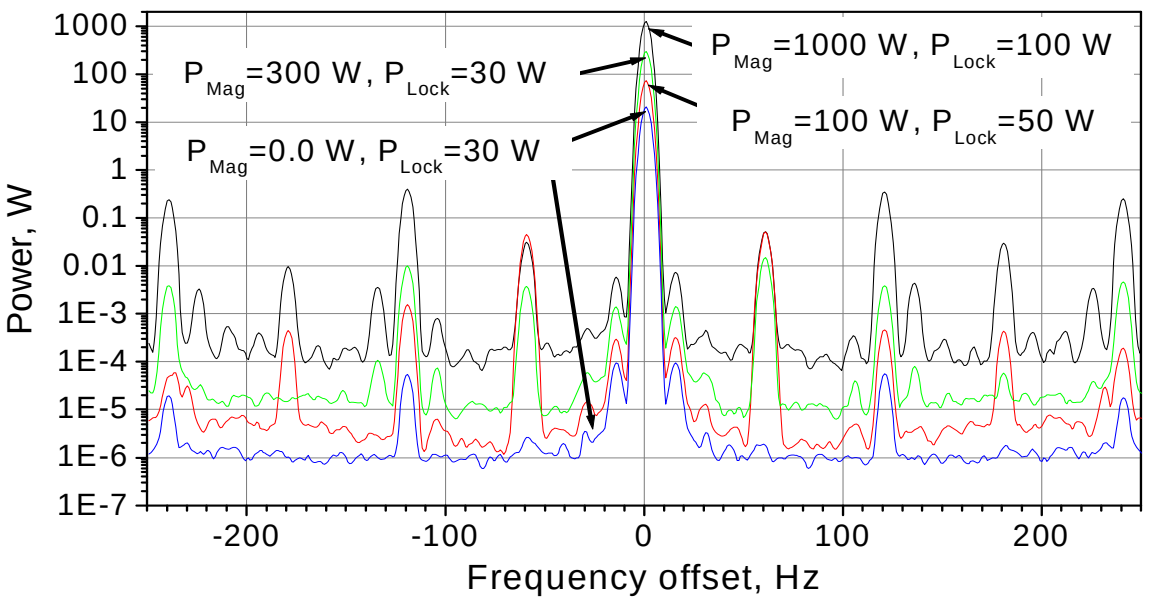

Figure 5: Offset of the carrier frequency at various powers of the magnetron, $P_{Mag}$, and the locking signal, $P_{Lock}$, [6].

The trace $P_{Mag}$ =0.0 W, $P_{Lock}$ =30 W, in Fig. 5 was measured with the magnetron high voltage turned OFF. It shows the frequency offset of the injection-locking signal.

Measured magnetron average efficiency vs. the range of power regulations at various methods of power control is plotted in Fig. 6, [6].

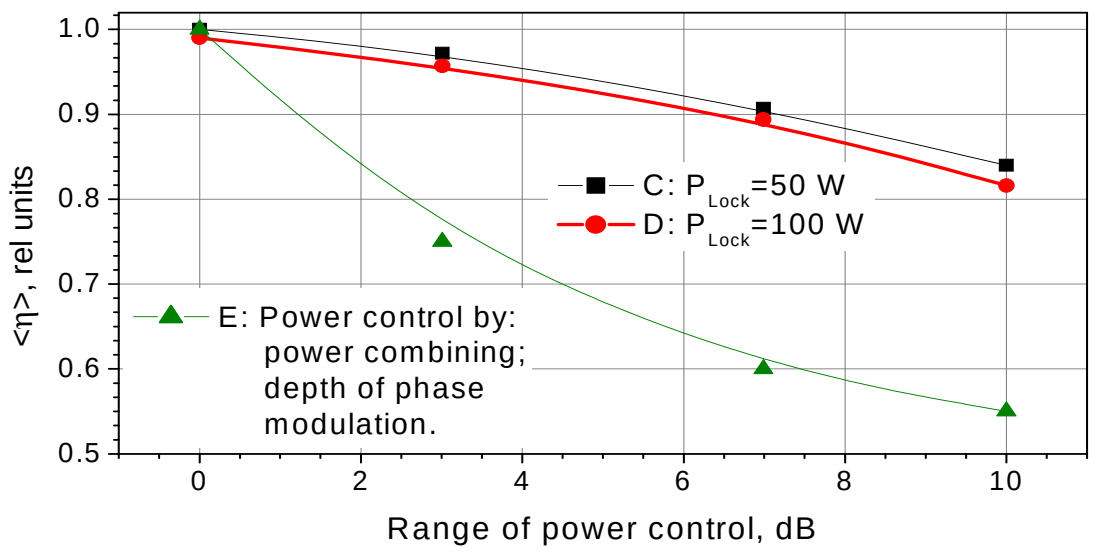

Figure 6: Relative magnetron efficiency vs. range of power variations at various methods of control. Traces C and D are average efficiency of the 1.2 kW magnetron, at the extended current control, [6]. Trace E is the average efficiency of 1 kW magnetrons at power combining, [4], or at management of the depth of phase modulation, [5].

Dynamic power control using this technique was demonstrated by modulating the magnetron current via the HV power supply. Fig. 7. It indicates that the averaged over 16 traces the magnetron output power tracks the harmonic modulations of the magnetron current.

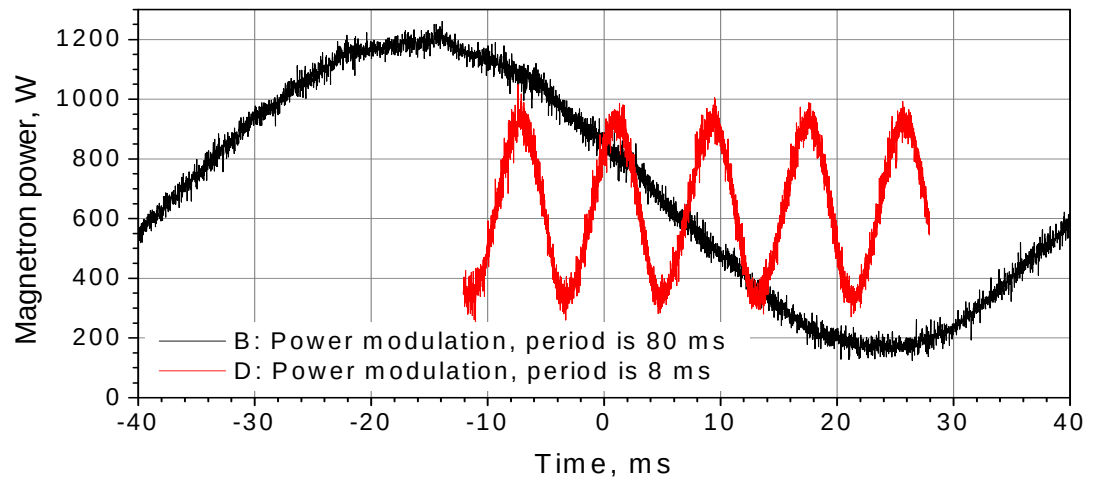

Figure 7: Modulation of the magnetron power managing the magnetron current by a harmonic signal controlling the SM445G HV switching power supply, [6].

## A HIGHLY-EFFICIENT MAGNETRON TRANSMITTER

A highly-efficiency magnetron transmitter based on a 2-cascade magnetron configuration described above is shown in Fig. 8, [7]. Phase is controlled by the injection-locking phase-modulated signal. Power control is realized by modulation of the low-power HV power supply biased by an uncontrolled main magnetron HV power supply within a fast current feedback loop. Power of the controlled HV power supply does not exceed 10% of power of the main power supply. Elimination of the phase pushing in the high-power magnetron caused by current regulation to the level of -50 dB or less provides a wide-band phase control by a Low Level RF (LLRF) system within a fast phase feedback loop. The bandwidth of power control up to 10 kHz is sufficient to achieve the rms standard deviations of the accelerating field amplitude in SRF cavity <1% at the beam current ≥5 mA.

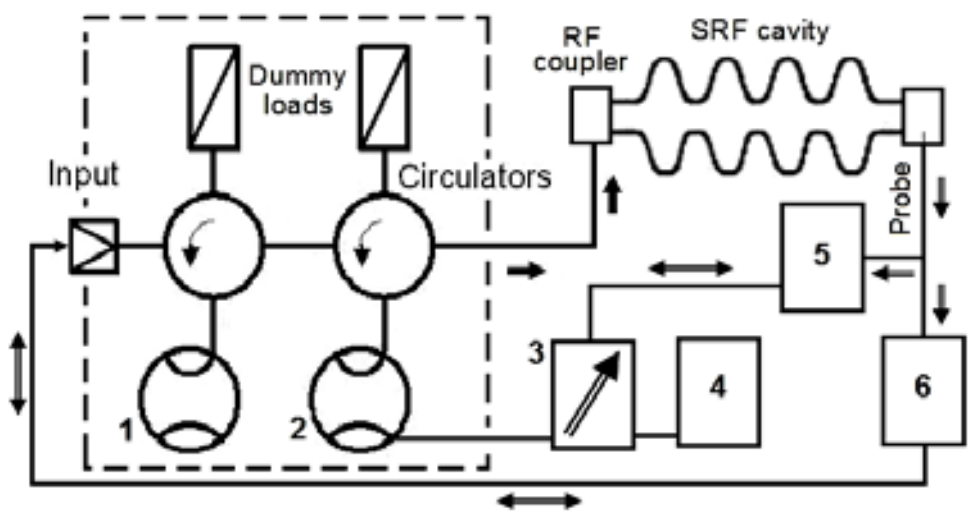

Figure 8: Conceptual scheme of the 2-cascade magnetron system for wide-band phase and mid-frequency power control over a 10 dB range in the SRF cavity of ADS-class accelerators. 1- the low power magnetron, 2- the high-power magnetron, 3- the low-power HV power supply controlled within the feedback loop, 4- the main uncontrolled HV power supply, 5- the current/voltage controller within the LLRF system for the low-power supply, 6- the phase controller within the LLRF system.

## SUMMARY

The experiments with 2.45 GHz, 1 kW, CW tubes demonstrated capabilities of the magnetrons controlled by a phase-modulated injection-locking signal for dynamic phase control required for intensity-frontier superconducting accelerators. Vector methods of power control in magnetrons are equally fast, but provide limited average efficiency at the control. The novel method of power control via wide-range control of the magnetron current provides highest efficiency at low noise and can minimizing the capital and operating costs for ADS-class projects.

## REFERENCES

[1] S. Simrock et al., in Proceed. of PAC 2003, 470-472, 2003.

[2] Z. Conwey, and M. Liepe, TU5PFP043, in Proceed. of PAC09, 1-3, 2009.

[3] G. Kazakevich, EIC 2014, TUDF1132_Talk, http://appora.fnal.gov/pls/eic14/agenda.full

[4] G. Kazakevich et al., NIM A 760 (2014) 19–27.

[5] B. Chase et al., JINST, 10, P03007, 2015.

[6] G. Kazakevich et al., NIM A 839 (2016) 43-51.

[7] G. Kazakevich et al., “A novel Technique of Power Control in Magnetron Transmitters for intense Accele Rators”, in Proceed. of NA-PAC 16, 2016.